\documentclass[
	physrev,
	reprint,
	amsmath,
	amsfonts,
	]{revtex4-2}

\usepackage[all]{nowidow}
\usepackage{csquotes}

\newcommand{\EQ}[3]{
  \begin{equation}
    \label{#1}
    #2
    \;#3
  \end{equation}
}

\usepackage{mathtools}

\usepackage[detect-all]{siunitx}
\usepackage{dcolumn}

\begin{document}

\title{Time-dependent condensate fraction in an analytical model}

\author{A. Simon}
%\email{hoelck@thphys.uni-heidelberg.de}
\author{G. Wolschin}
\email{wolschin@thphys.uni-heidelberg.de}
\affiliation{Institute for Theoretical Physics, Heidelberg University, Philosophenweg 12--16, D-69120 Heidelberg, Germany, European Union}

\date{\today}

\begin{abstract}
We apply analytical solutions of a nonlinear boson diffusion equation (NBDE) that include boundary conditions at the singularity
to calculate the time evolution of the entropy during evaporative cooling
of ultracold atoms, and the time-dependent condensate fraction. For suitable initial conditions it is found to agree with available data on $^{23}$Na.\\
\end{abstract}

\maketitle

\section{Introduction}
%\label{}
Shortly after the discovery of Bose-Einstein condensation in ultracold atoms of rubidium \cite{an95}, sodium \cite{ket95} and lithium \cite{hul95,hul97}, models and theories to calculate the time-dependent condensate fraction were developed, see Refs.\,\cite{gz97,gardiner_quantum_1997,bzs00} and related works. These approaches usually consider a quantum treatment such as  a nonlinear Schr\"odinger equation for the condensate and an equilibrium- or nonequilibrium-statistical description for the thermal component. 

Discrepancies between early theoretical results \cite{gz97} and data for the time-dependent condensate fraction \cite{miesner_bosonic_1998} were partly due to the fact that the noncondensed particles were represented by an equilibrium particle reservoir with a fixed chemical potential -- an assumption that does not correspond to the nonequilibrium initial state that is generated through evaporative cooling. 

The relaxational dynamics of the nonthermal component that occurs in the presence of the mean field of the condensate was later taken into account through kinetic equations. In particular, a semiclassical Boltzmann equation \cite{eckern84,kd85,sto97,stoof_coherent_1999,zng99,bzs00} was solved numerically in the ergodic approximation, which has been used by many authors \cite{snowo89,kss92,setk95,lrw96,hwc97,jgz97}. In this approximation, the population of a given state depends only on its energy. Such an approach provides a detailed description of the growth of the condensate based on numerical solutions of the equations for the dynamics of the thermal cloud and its mean-field interaction with the condensate \cite{bzs00}, yielding satisfactory -- though not perfect -- agreement with the available data on the time-dependent condensate fraction.
These numerical approaches to thermalization and condensate formation usually rely on an artificial seed condensate \cite{bzs00}. 

In view of such involved numerical approaches it is interesting to have a simple, exactly solvable
model for the physics of condensate formation in cold atoms. For bosons, such a model has been proposed in Ref.\,\cite{gw18} and adapted to cold quantum
gases in Refs.\,\cite{gw18a,gw20}. The model is based on a nonlinear boson diffusion equation (NBDE) which has been derived from the quantum Boltzmann collision term. For energy-independent
transport coefficients, it is structurally simple, but still complicated to solve exactly due to the nonlinearity in the drift term, which corresponds to Bose stimulation and causes the system to reach the Bose-Einstein equilibrium distribution for sufficiently large times, with a singularity in the infrared. The solutions represent, in particular, evaporative cooling from an initial temperature $T_\text{i}$ to a final temperature $T_\text{f}$ that may be below the critical value $T_\text{c}$. We explore the solutions in this work to calculate the time-dependent condensate fraction based on the analytic solutions and on particle-number conservation in the cloud and the condensate, together with the time evolution of the entropy in successive and single cooling steps.

%With these exact solutions, the time-dependent entropy of an equilibrating finite Bose-system has been calculated in Ref.\cite{gw20}.
In the next section, the NBDE and its exact solutions \cite{gw18a,gw20}  with the consideration of boundary conditions at the singularity are briefly reviewed. In Section\,3, the analytic time-dependent solutions of the nonlinear equation with constant transport coefficients are evaluated and compared to numerical results. In Section\,4, the time evolution of the entropy in evaporative cooling is considered for several sequential cooling steps, as well as for a single step that leads to condensate formation.  In Section\,5, the time-dependent chemical potential is evaluated from the condition of particle-number conservation and the condensate fraction is discussed based on the analytic NBDE solutions. Due to the quantum-statistical properties of the bosonic system that are encoded in the NBDE, a condensate forms and grows if the final temperature is below the critical value. With parameters adapted to the evaporative cooling of $^{23}$Na, a comparison of the time-dependent condensate fraction with MIT data is presented. The conclusions are drawn in Section\,6.
% which are also compared to the linear relaxation ansatz. 

\section{Nonlinear boson diffusion equation and analytic solutions}
The system of ultracold atoms is viewed as a time-dependent mean-field supplemented by a collision term. The atomic density is usually  so low that interactions involving more than two particles at a time can be ignored. Only $s$-wave scattering contributes for bosons at ultracold temperatures, such that the interaction is a contact interaction with a coupling strength that is proportional to the $s$-wave scattering length.

The $N$-body density operator $\hat{\rho}_N(t)$ can be written based on $N$ single-particle wave functions of the atoms which are solutions of the time-dependent Hartree-Fock equations supplemented by a time-irreversible collision term $K_N(t)$ that accounts for the thermalization through random two-body collisions  ($\hbar=c=1$)
\EQ{}{
i\,\frac{ \partial\hat{\rho}_N(t)}{\partial t} = \big[\hat{H}_\text{HF}(t),\hat{\rho}_N(t)\big] + i \hat{K}_N(t)
\label{eq1}}{}
where $\hat{H}_\text{HF}(t)$ is the self-consistent Hartree-Fock mean field of the atoms, with an external potential for ultracold atoms in a trap.
 
In this work we do not consider the full many-body problem and its reduction to the one-body level, but rather an approximate version that starts from the reduced ensemble-averaged single-particle density operator $\bar{\rho}_1(t)$. Its diagonal elements can be interpreted as the 
probability for a particle to be in a state $|\alpha\rangle$ with energy $\epsilon_\alpha$
\EQ{}
{\big(\bar{\rho}_1(t)\big)_{\alpha,\alpha} = \langle n(\epsilon_\alpha, t)\rangle \equiv n_\alpha(\epsilon,t)}
{.}
The total number of particles is $N=\sum_\alpha n_\alpha$, and we neglect here
the off-diagonal terms of the density matrix. The occupation-number distribution $n_\alpha(t)$ in a finite Bose system obeys a Boltzmann-like collision term \cite{gw18}
\begin{align}
\frac{\partial n_\alpha}{\partial t }= \sum_{\beta,\gamma,\delta} \overline{V^2_{\alpha\beta\gamma\delta}}\,\mathcal{G}_{\alpha\beta\gamma\delta} \big[ (1+n_\alpha)(1+n_\beta)n_\gamma n_\delta\\ \nonumber
-(1+n_\gamma)(1+n_\delta)n_\alpha n_\beta \big]
\end{align}
where $\overline{V^2_{\alpha\beta\gamma\delta}}$ is the second moment of 
the pairwise atom-atom interaction. $\mathcal{G}_{\alpha\beta\gamma\delta}$ is an energy-conserving function which has a finite width, because it is the total energy of mean field plus collision term that must be conserved.  It becomes a $\delta$-function as in the Boltzmann-Nordheim equation \cite{no28} only if the mean-field energy content stays constant in time -- which is, however, usually not the case in cold-atom systems. Whereas in an infinite homogeneous system that is described using a quantum Boltzmann
equation with an energy-conserving $\delta$-function an infinite amount of time is required to create a condensate, this is not expected to be the case for an energy-conserving function that has a finite width. Hence, the theoretical framework is consistent with a finite condensate formation time as required from experiment.

This equation has been transformed into a nonlinear partial differential equation in Refs.\,\cite{gw18,gw18a}. One starts by defining transition probabilities from state $\gamma$ to $\alpha$ as
\EQ{}{
W_{\gamma \rightarrow \alpha} = \sum_{\beta,\delta} \overline{V^2_{\alpha\beta\gamma\delta}} \,\mathcal{G}_{\alpha\beta\gamma\delta} (1+n_\beta)\,n_\delta
}{.}	
The collision term can be written as a
master equation where the transitions into and out of the state with energy $\epsilon_\alpha$ are more explicit
\EQ{}
{
\frac{\partial n_\alpha}{\partial t }= (1+n_\alpha) \sum_\gamma W_{\gamma \rightarrow \alpha} n_\gamma 
- n_\alpha \sum_\gamma W_{\alpha \rightarrow \gamma} (1+n_\gamma)
}{.}
Introducing the density of states  $g_\alpha = g(\epsilon_\alpha)$, such that $W_{\gamma \rightarrow \alpha} = W_{\gamma,\alpha} g_\alpha$ 
and using the fact that quantum particles are indistinguishable
$W_{\alpha,\gamma} = W_{\gamma,\alpha} = W(\epsilon_\alpha, \epsilon_\gamma)$ one arrives at 
\EQ{}
{
\frac{\partial n_\alpha}{\partial t }=  \int_0^\infty W_{\alpha,\gamma} \left[ g_\alpha (1+n_\alpha) n_\gamma - g_\gamma (1+n_\gamma) n_\alpha \right] \text{d} \epsilon_\gamma
}{.}
    Defining the transport coefficients $v\equiv v\,(\epsilon_\alpha,t)$ and $D\equiv D\,(\epsilon_\alpha,t)$ as first and second moments of the transition probability \cite{gw18},
%%\begin{align}
%D &\equiv D\,(\epsilon_\alpha,t) =  \frac{1}{2} g_\alpha \int_0^\infty W_{\alpha,\gamma} (\epsilon_\gamma -\epsilon_\alpha)^2 \text{d} \epsilon_\gamma\,,\\
%v &\equiv v\,(\epsilon_\alpha,t) = g_\alpha^{-1} \frac{\partial}{\partial \epsilon_\alpha}{\left[\,g_\alpha D\,\right]}\,,
%\end{align}
we have obtained the nonlinear boson diffusion equation (NBDE) for the single-particle expectation-value occupation-number distribution of the energy eigenstates $\epsilon_\alpha$, $n\equiv n_\alpha \equiv \langle n(\epsilon_\alpha,t)\rangle$, in Refs.\,\cite{gw18,gw18a} as \footnote{The derivative-term of the diffusion coefficient has been modified as compared to Refs.\,\cite{gw18,gw18a,gw20} in order to secure the correct stationary solution}
 \begin{equation}
\frac{\partial n}{\partial t}=-\frac{\partial}{\partial\epsilon}\Bigl[v\,n\,(1+n)+n\frac{\partial D}{\partial \epsilon}\Bigr]+\frac{\partial^2}{\partial\epsilon^2}\bigl[D\,n\bigr]\,.
 \label{boseq}
\end{equation}
The drift term $v(\epsilon,t)$ accounts for dissipative effects, the term $D(\epsilon,t)$ for diffusion of particles in the energy space.
The many-body physics is contained in these
transport coefficients, which depend on energy, time, and the second moment of the interaction.

A prerequisite for the reduction to $1+1$ dimensions in the above formulation is spatial and momentum isotropy. This corresponds to the assumption of sufficient ergodicity, which has been widely discussed in the literature  \cite{snowo89,kss92,setk95,lrw96,hwc97,jgz97}. For the thermal cloud of cold atoms around a Bose-Einstein condensate (BEC), it is expected to be a reasonable assumption, even though the condensate in a trap is spatially anisotropic. Regarding the role of different spatial dimensions in view of BEC formation, this enters our present formulation only through the density of states, which differs according to the number of spatial dimensions, and the type of confinement. The model calculations in this work are for a 3d system, and we shall investigate results for the density of states of a free Bose gas, and bosonic atoms confined in a harmonic trap. One-dimensional systems where no BEC should be formed have not yet been studied.

To derive the stationary solution $n_\infty(\epsilon)$ for the equation with
variable transport coefficients, we rewrite Eq.\,(\ref{eq1}) and set the time derivative
to zero
\begin{equation}
0= \frac{\partial}{\partial \epsilon}{\left[ v\,n_\infty \,(1+n_\infty)- D\,\frac{\partial n_\infty}{\partial \epsilon} \right]}\,,
\end{equation}
such that
\EQ{}
{v\,n_\infty(1+n_\infty) - D\,\frac{\partial n_\infty}{\partial \epsilon} = c_1}{.}
Dividing by $n_\infty(1+n_\infty)D$ and integrating over $\epsilon$ yields
\EQ{}
{
\int \frac{\text{d} n_\infty/\text{d} \epsilon}{n_\infty(1+n_\infty)}\text{d}\epsilon = 
\int \left( \frac{v}{D} - \frac{c_1}{n_\infty(1+n_\infty)D} \right)\text{d} \epsilon 
}{.}
We integrate the l.h.s, resulting
in $[\ln(n_\infty) - \ln(1+n_\infty)] = \ln(1-\frac{1}{1+n_\infty})$ plus an integration constant $c_2$.
Solving for $n_\infty$ one obtains
\EQ{}
{n_\infty = \left[
\exp\left( 
\int  \left(- \frac{v}{D} + \frac{c_1}{n_\infty(1+n_\infty)D}\right)\text{d} \epsilon +c_2 \right) -1
\right]^{-1}.
}{}
%The stationary solution $n_\infty$ is independent of time, and we set $c_1(t)\equiv 0~ \forall \,t$.
In order to reduce to a Bose-Einstein distribution, $c_1 = 0$ is required, and the ratio $v/D$ must have no energy dependence for ${t\rightarrow\infty}$ so that it can 
be pulled out of the integral.
% The remaining integral can then be integrated to give a factor $\epsilon$. 
It follows that $\lim \limits_{t \to \infty}[-v(\epsilon,t)/D(\epsilon,t)] \equiv 1/T$ and $c_2\equiv -\mu/T$ such that the stationary distribution equals the Bose-Einstein equilibrium distribution ($k_\text{B}=1$)
\begin{equation}
n_\infty(\epsilon)=n_\text{eq}(\epsilon)=\frac{1}{e^{(\epsilon-\mu)/T}-1}
 \label{Bose-Einstein}
\end{equation}
with the chemical potential $\mu<0$ in a finite Bose system.
%This result implies also that the equation with constant transport coefficients \eqref{eq:const_NBDE} has a Bose-Einstein distribution as
%its stationary solution. 

In the limit of energy-independent transport coefficients the nonlinear boson diffusion equation  
for the occupation-number expectation-value distribution $n(\epsilon,t)$
% per unit volume  
becomes
%has the simple form
\begin{equation}
\frac{\partial n}{\partial t}=-v\,\frac{\partial}{\partial\epsilon}\Bigl[n\,(1+n)\Bigr]+D\,\frac{\partial^2n}{\partial\epsilon^2}\,.
 \label{bose}
\end{equation}
As in case of Eq.\,(\ref{boseq}), the thermal equilibrium distribution $n_\text{eq}$ is a stationary solution
with $\mu<0$ and $T=-D/v$. 

In spite of its simple structure, the NBDE with constant transport coefficients thus
preserves the essential features of Bose-Einstein
statistics which are contained in the bosonic Boltzmann equation. For a given initial condition $n_\text{i}(\epsilon)$, it can be solved exactly using the nonlinear transformation outlined
in Ref.\,\cite{gw18a}. The resulting solution is
%For an initial distribution \( n_{\mathrm{i}}(\epsilon) \), the solution for the time-dependent occupation-number distribution becomes
\begin{align}
    n(\epsilon,t) = -\frac{D}{v} \frac{\partial}{\partial\epsilon}\ln{\mathcal{Z}(\epsilon,t)} -\frac{1}{2}= -\frac{D}{v}\frac{1}{\mathcal{Z}} \frac{\partial\mathcal{Z}}{\partial\epsilon} -\frac{1}{2}
    \label{eq:Nformula} 
    \end{align}
 where the time-dependent partition function ${\mathcal{Z}(\epsilon,t)}$ obeys a linear diffusion equation
     \begin{align}
    \frac{\partial}{\partial t}{\mathcal{Z}}(\epsilon,t) = D \frac{\partial^2}{\partial\epsilon^2}{\mathcal{Z}}(\epsilon,t)\,.
    \label{eq:diffusionequation}
\end{align}
The partition function $\mathcal{Z}(\epsilon,t)$
     \begin{align}
    \mathcal{Z}(\epsilon,t)= a(t)\int_{-\infty}^{+\infty} G(\epsilon,x,t)\,F(x)\,\text{d}x
    \label{eq:partitionfunctionZ}
    \end{align}
 is an integral over Green's function $G(\epsilon,x,t)/\sqrt{4\pi\,Dt}$ of Eq.\,(\ref{eq:diffusionequation}), and an exponential function $F(x)$ that depends on 
 %an integral over 
 the initial occupation-number distribution $n_\text{i}$
\begin{align}
    F(x) = \exp\Bigl[ -\frac{1}{2D}\bigl( v x+2v \int_0^x n_{\mathrm{i}}(y)\,\text{d}y \bigr) \Bigr]\,.
       \label{ini}
\end{align}
The time-dependent prefactor $a(t)=1/\sqrt{4\pi\,Dt}$ in Eq.\,(\ref{eq:partitionfunctionZ}) cancels out when taking the logarithmic derivative in Eq.\,(\ref{eq:Nformula}), and for the same reason the
 definite integral over the initial conditions taken at the lower limit in Eq.\,(\ref{ini}) drops out in the calculation of $n(\epsilon,t)$. Hence, the definite integral in Eq.\,(\ref{ini}) can be replaced \cite{rgw20} by the indefinite integral $A_{\mathrm{i}}(x)$ over the initial distribution with $\partial_x A_{\mathrm{i}} (x) = n_{\mathrm{i}}(y)$.
 %, such that
%\begin{align}
%    F(x) = \exp\Bigl[-\frac{1}{2D}\left( v x+2v A_{\mathrm{i}}(x) \right)\Bigr]\,.
%    \label{eq:G(x)}
%\end{align}
%This replacement still provides the exact solution. 

With these modifications that do not affect the accuracy, it becomes possible to compute the partition function and the overall solution for the occupation-number distribution function Eq.\,(\ref{eq:Nformula}) analytically.
%, even in the presence of a singularity in the initial conditions, and with boundary conditions at the singularity $\epsilon = \mu < 0$ \cite{gw20}.

When solving the problem with the free Green's function of Eq.\,(\ref{eq:diffusionequation}), the physically correct solution with the Bose-Einstein equilibrium limit is attained in the UV region, but not in the IR \cite{gw18}. To solve this problem, one has to consider the boundary conditions at the singularity $\epsilon = \mu < 0$ \cite{gw20}. 
For fixed chemical potential $\mu$, an initial temperature $T_\text{i}$, and a final temperature $T_\text{f} = - D/v$, the combined initial- and boundary value problem has been solved exactly in Refs.\,\cite{rgw20,gw20} using the above nonlinear transformation from Eq.\,(\ref{eq:Nformula}) \cite{gw18,gw20}, and an infinite series expansion. The analytic solutions are briefly reconsidered before we use them for further calculations and in particular, to obtain the time-dependent condensate fraction.

%\section{Exact solution with boundary conditions}
%To solve the problem with boundary conditions at the singularity,  
%$n(\epsilon=\mu,t)=\infty$, 
%the chemical potential is first treated as a fixed parameter. This will provide us with an exact solution -- which, however, violates particle-number conservation: To preserve the particle %number in the course of the time evolution, the chemical potential must approach zero, in accordance with the fact that atoms move into the condensed state. 

With the boundary condition \(\lim_{\epsilon \downarrow \mu} n(\epsilon,t) = \infty\) \,$\forall$ \(t\) at the singularity $\epsilon = \mu$  one obtains \( \mathcal{Z} (\mu,t) = 0\), and the energy range is restricted to  $\epsilon \ge \mu$. This requires a new Green's function \cite{gw20} 
%\( {F} (\epsilon,x,t) \) \cite{EqWorld} 
that equals zero at \(\epsilon = \mu\) $\forall \,t$. It can be written as
\begin{align}
    {G} (\epsilon,x,t) = G_\text{free}(\epsilon - \mu,x,t) - G_\text{free}(\epsilon - \mu,-x,t)\,,
    \label{eq:newGreens}
\end{align}
and the partition function with this boundary condition becomes
\begin{align}
    {\mathcal{Z}} (\epsilon,t) = \int_0^\infty {G} (\epsilon, x, t)\,F(x+\mu)\, \text{d}x\,.
    \label{eq:newformulaforZ}
\end{align}
%Here, the Green's function Eq.\,\eqref{eq:newGreens} restricts the integral to energies $\epsilon \ge \mu$. 
%This gives rise to a new expression for the occupation-number distribution.
%The corresponding solution for the occupation-number distribution with fixed chemical potential is equivalent to imposing a point symmetry around \(\mu\) in the initial distribution $n_\text{i}(\epsilon)$ that appears in \(G(x)\).
The function $F$ remains unaltered with respect to Eq.\,(\ref{ini}), but its argument is shifted by the chemical potential.

\begin{figure}[b!]
    \centering
    \includegraphics[width = 0.47\textwidth]{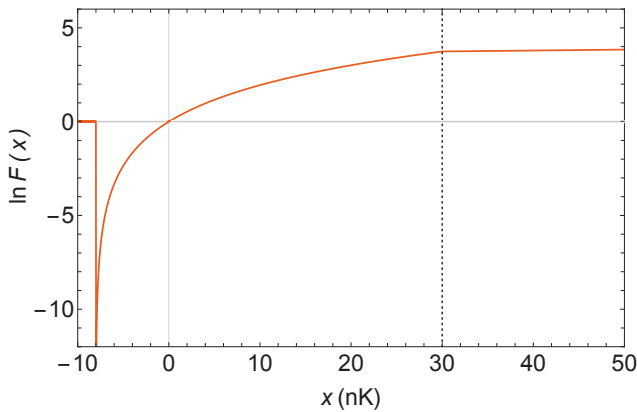}
    \caption{The argument $\ln{F(x)}$ of the exponential function $F(x)$ in Eq.\,(\ref{ini}) (solid curve) with a singularity at  \(x = \mu<0\) in the definite integral of an initial distribution $n_\text{i}$ given by a Bose-Einstein distribution with $T=240$\,nK and $\mu=-8$\,nK that is truncated at $30$ nK (dotted vertical line).     %The parameters are as in Ref.\,\cite{gw18a}, and in Section\,III\,C. The dashed curve is the argument 
   % $\ln{F(x)}$ for the corresponding indefinite integral of the initial distribution (primitive) $A_\text{i}(x)$ with an integration constant $c\equiv 0$, see text. }
   } 
    \label{fig1}
\end{figure}

As an initial condition that is appropriate for a schematic description of evaporative cooling {\cite{an95,dmk95,lrw96}} to demonstrate our method, a truncated thermal equilibrium distribution that is cut off at a maximum energy $\epsilon_\text{i}$ beyond which high-velocity atoms are removed has been chosen in our previous works 
\begin{equation}
n_\text{i}(\epsilon)=\frac{1}{e^{(\epsilon-\mu)/T}-1}\, \theta (1-\epsilon/\epsilon_\text{i})\,.
 \label{inibec}
\end{equation}
The distribution is truncated at a maximum energy $\epsilon_\text{i}$ beyond which high-velocity atoms are removed. In a trapped ultracold gas, these atoms leave the trap: The system of atoms in the trap is thus not isolated  and its entropy decreases in the course of cooling. In the subsequent re-thermalization (and for $T_\text{f}<T_\text{c}$, condensate formation), the entropy of the atoms in the trap rises again, but remains below the initial thermal value. This will be discussed in more detail in Section\,5.

%The argument $\ln[F(x)]$ of the exponential function with a singularity at $x=\mu$ (before shifting it) is shown in  Fig.\,\ref{fig1} for a schematic set of parameters: $T_\text{i} = 240$ nK,
For such a truncated initial thermal distribution, the 
 logarithm of the function $F(x)$ with a singularity at {$x=\mu$}\\ (before shifting it by $\mu$) is shown in  Fig.\,\ref{fig1} with the following set of parameters: $T_\text{i} = 240$ nK,
 $\epsilon_\text{i} = 30$ nK, $D = 100\,$(nK)$^2$\,ms$^{-1}$, $v = -1\,$nK\,ms$^{-1}$, $T_\text{f} = - D/v = 100$ nK. 
 %These values are roughly appropriate for evaporative cooling and condensation of $^{87}$Rb. 
 Due to the singularity, $F(x)$ vanishes at $x = \mu=-8$\,nK. The logarithm of $F(x)$ is continuous, but not differentiable at $x = \epsilon_\text{i}$, which holds the key to the equilibration in the UV region.
% $F(x)\rightarrow F(x+\mu)$.

Using Eq.\,(\ref{inibec}) as initial distribution, the occupation-number distribution 
%Eq.\,(\ref{eq:Nformula})
%with boundary conditions at the singularity 
can be evaluated exactly for $T_\text{i} \ne T_\text{f}$ with the time-dependent partition function \cite{rgw20}
%as shown in the next section. The result for constant $T$ \cite{rgw20} 
%For the same initial distribution Eq.\,(\ref{inibec}), the partition function $ {\mathcal{Z}}$ can thus be evaluated with boundary conditions using ${G} (\epsilon, x, t)$ as %Green's function, and replacing $F(x)\rightarrow F(x+\mu)$.
%The occupation-number distribution function Eq.\,\ref{eq:Nformula} with boundary conditions at the singularity and $\mathcal{Z}\rightarrow {\mathcal{Z}}$ can still be evaluated %exactly. The result \cite{rgw20} 
%\begin{align}
%    {n}(\epsilon,t) = \frac{1}{\exp\Bigl(\frac{\epsilon-\mu}{T}\Bigr) L(\epsilon,t)-1}
%    \label{eq:particledistributionfixedmu}
%\end{align}
%is formally similar to a Bose-Einstein distribution. In the next section, different initial and final temperatures $T_\text{i} \ne T_\text{f}$ are considered to schematically account for %evaporative cooling.
%	\section{Analytical solutions for cooling}

%	If the temperature $T_\text{i}$ in the initial conditions Eq.\;(\ref{inibec}) differs from the final equilibrium temperature $T_\text{f}=-D/v$ as is the case in evaporative cooling  %{\cite{an95,dmk95,lrw96}}, the analytic solutions of the NBDE with boundary conditions become more involved, but it is still possible to derive them. 
	%The analytical solutions of the nonlinear NBDE produce a realistic account of the thermalization. 
%	The partition function with boundary conditions at the singularity 
%$\epsilon=\mu$ has been derived in Ref.\, \cite{rgw20} for $T_\text{i}\ne T_\text{f}$ as\\

\begin{align}
{\mathcal{Z}}(\epsilon,t) = \sqrt{4 D t} \, \exp\Bigl(-\frac{\mu}{2 T_{\mathrm{f}}}\Bigr) \sum_{k=0}^{\infty} \binom{\frac{T_{\mathrm{i}}}{T_{\mathrm{f}}}}{k} \left( -1 \right)^k \times  \notag \\
    \Bigg( \text{e}^{\alpha_k^2 D t} \left[ \text{e}^{\alpha_k (\epsilon - \mu)} \Lambda_1^k (\epsilon,t) - \text{e}^{\alpha_k (\mu - \epsilon)} \Lambda_2^k (\epsilon, t) \right] \notag \\
    + \exp\Bigl( \frac{(\mu - \epsilon_i)k}{T_{\mathrm{i}}}\Bigr) \exp\Bigl({\frac{D t}{4 T_{\mathrm{f}}^2}}\Bigr) \times \notag \\
  \Big[ \exp\Bigl(\frac{\epsilon-\mu}{2 T_{\mathrm{f}}}\Bigr) \Lambda_3 (\epsilon,t) - \exp\Bigl(\frac{\mu - \epsilon}{2 T_{\mathrm{f}}}\Bigr) \Lambda_4 (\epsilon,t) \Big] \Bigg)\,    \label{eq:Zarbtemp}
\end{align}
where $\alpha_k=1/T_\text{f}-k/T_\text{i}$ and the auxiliary functions are
\begin{align}
    \Lambda_1^{k} (\epsilon,t) =& \,\text{erf}\,\Bigl(\frac{\epsilon - \mu +2 D t \alpha_k}{\sqrt{4 D t}}\Bigr) \notag  \\
        &\qquad -  \,\text{erf}\,\Bigl(\frac{\epsilon - \epsilon_i + 2 D t \alpha_k}{\sqrt{4 D t}}\Bigr)\,,\\
    \Lambda_2^{k} (\epsilon,t) =& \,\text{erf}\,\Bigl(\frac{\mu-\epsilon+ 2 D t \alpha_k}{\sqrt{4 D t}}\Bigr) \notag \\ 
    &\qquad -  \,\text{erf}\,\Bigl(\frac{2 \mu - \epsilon - \epsilon_i + 2 D t \alpha_k}{\sqrt{4 D t}}\Bigr)\,,\\
    \Lambda_3 ( \epsilon,t) =& \,\text{erfc}\,\Bigl(\frac{\epsilon_i - \epsilon + t v }{\sqrt{4 D t}}\Bigr)\,, \\
    \Lambda_4 ( \epsilon,t) =& \,\text{erfc}\,\Bigl(\frac{\epsilon - 2 \mu + \epsilon_i + t v}{\sqrt{4 D t}}\Bigr)\,.
\end{align}

	 	\begin{figure}[t!]
	\centering
	\includegraphics[width = 0.47\textwidth]{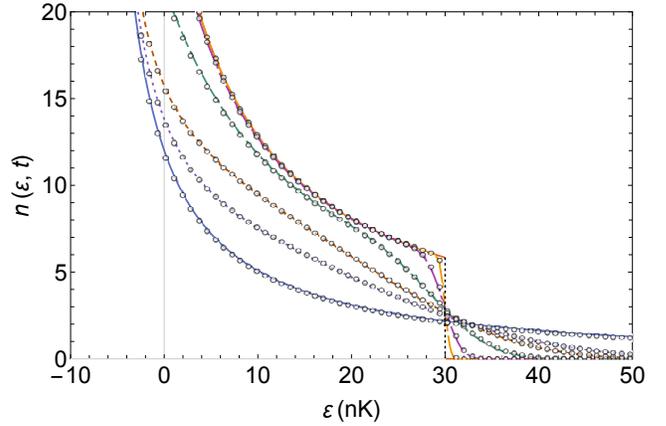}
	\caption{\label{fig2}  Schematic representation of evaporative cooling in a a finite Bose system based on the nonlinear evolution according to Eq.\,(\ref{bose})
	starting from a truncated Bose-Einstein distribution Eq.\,(\ref{inibec}), upper solid curve with cutoff at $\epsilon_\text{i}$\,=\,30\,nK and $\mu=-8$\,nK.
	The transport coefficients are $D = 100$\,(nK)$^2$\,ms$^{-1},~\,v = -1$\,nK$\,$ms$^{-1}$. The initial temperature is $T = 240$\,nK, the final temperature 
	$ T_\text{f} = T = -D/v = 100$\,nK.
	The time evolution of the single-particle occupation-number distributions is shown at $t = 0.001, 0.01, 0.1,  0.4$, and $0.8$\, ms (top to bottom in the IR). For comparison, the numerical results using Matlab are shown as open circles.
%It recovers the equilibrium distribution with temperature $T$ at $t\simeq 40$ ms.
	}
	\end{figure}
	
To obtain the time-dependent occupation-number distribution function from Eq.\,(\ref{eq:Nformula}), the derivative $\partial  {\mathcal{Z}}/\partial\epsilon$ is also required. It can be calculated analytically \cite{rgw20}, such that an exact expression for the time-dependent occupation-number distribution function for evaporative cooling results that can be directly compared with numerical solutions of the NBDE, see the following section.%For $T_\text{i}=T_\text{f}$, the result of Eq.\,(\ref{eq:particledistributionfixedmu}) is recovered.
\section{Occupation-number distributions}
The time-dependent analytic distribution functions that solve the NBDE exactly with boundary conditions at the singularity are displayed in Fig.\;\ref{fig2} for the same initial conditions as in Fig.\;\ref{fig1}. The analytic results agree precisely with numerical solutions (open circles) of the basic equation,  {and predict the time-dependent cooling from a thermal distribution with temperature $T_\text{i}$ that is truncated at $\epsilon_\text{i}$ to a Bose-Einstein distribution with $T_\text{f} < T_\text{i}$, which is the thermal distribution for $t\rightarrow \infty$. The approach is similar to the kinetic theory of evaporative cooling in works such as Ref.\,\cite{lrw96}, but now an analytically solvable model is formulated.}

The corresponding numerical solutions (circles in Fig.\;\ref{fig2}) have been obtained using Matlab's routine \textit{pdepe}\rm\, for the solution of partial differential equations with given initial and boundary values \cite{skeel1990method}. Basic finite-difference algorithms like the Crank-Nicolson method did not provide sufficient accuracy for this nonlinear problem. 
The implementation of \textit{pdepe} is based on the algorithm described in Ref.~\cite{skeel1990method} and is suited for (nonlinear) parabolic 
partial differential equations. 
Due to the singularity at $\epsilon=\mu$, the integration was started at $\mu+\delta<0$. Regarding the associate boundary condition, we use the known stationary solution $n_\infty(\epsilon)$ to determine the boundary values in the IR as $n_\infty(\mu+\delta)$, and accordingly, in the UV. This minimizes the numerical errors at the boundaries. The value of $\delta$ is chosen as small as possible, but large enough to prevent numerical inaccuracies. In this particular case $\delta=0.2$ and an upper boundary of $\epsilon_\text{u}=3 \times T_\text{i}$ yields satisfactory agreement with the exact solutions.

 {The values of the transport coefficients $v, D$ in this specific model calculation have been derived from their relations to the equilibrium temperature $T=-D/v$ and the equilibration time $\tau_\text{eq}=4D/(9v^2)$ \cite{gw18}, with $T=100$ nK and $\tau_\text{eq}=44$ ms. They do not yet correspond to a specific experimental situation, and the cut-off temperature 
 $\epsilon_\text{i}=30$\,nK is articifially low to better demonstrate the behaviour of the analytic solutions. For direct comparisons with data, these values shall be adapted to the corresponding experiment, see Section\,5.}
% that can be obtained using Matlab. 

In the infrared, thermalization occurs faster \cite{gw20} than in case of a linear relaxation ansatz: With the parameter set of Fig.\;\ref{fig2}, the thermal distribution is reached within $t\simeq 1$ ms in the IR.  The buildup of the thermal slope in the UV is, however, slower in the nonlinear model as compared to the linear relaxation ansatz, because the latter enforces  {an exponential approach to {the} Boltzmann-like tail even for $\epsilon\rightarrow\infty$ \cite{gw20}.
%This will have a significant effect on the time-dependent entropy.
%\cite{bijlsma_condensate_2000}
	\begin{figure}[t!]
	\centering
	\includegraphics[width = 0.47\textwidth]{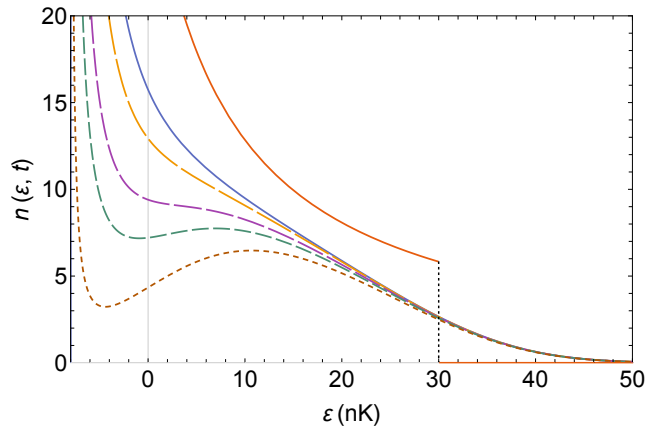}
	\caption{\label{fig3}  Convergence properties of the analytic NBDE solution with constant transport coefficients according to Eqs.\,(\ref{eq:Nformula})  and (\ref{eq:Zarbtemp}).
	The initial condition is as in Fig.\,\ref{fig2}, upper solid curve with cutoff at $\epsilon_\text{i}$\,=\,30\,nK.
	The time-dependent nonequilibrium solution is shown for $t = 0.4$\,ms, lower solid curve. The convergence is displayed for maximum expansion parameters
	$k_\text{max}=10, 12, 14, 20$, and $200$ (from bottom to top) in the series expansion of the partition function.
%It recovers the equilibrium distribution with temperature $T$ at $t\simeq 40$ ms.
	}
	\end{figure}

The convergence of our analytical method is displayed in Fig.\;\ref{fig3} using the above NBDE solution from Fig.\;\ref{fig2} for $t=0.4$\,ms as an example. Results according to Eq.\,(\ref{eq:Nformula}) are shown for $k_\text{max}=10, 12, 14, 20$, and $200$ (from bottom to top) in the series expansion of the partition function 
Eq.\,(\ref{eq:Zarbtemp}). Here, $k_\text{max}$ is the maximum expansion coefficient that is used in the calculation.  A solution for $k_\text{max}=40$ is already indistinguishable from the plotted one for $k_\text{max}=200$. It is thus sufficiently close to the exact solution, which would require $k_\text{max}=\infty$. We shall later check the convergence properties also for derived quantities such as the time-dependent chemical potential $\mu(t)$ that is needed to maintain particle-number conservation in the course of the time evolution, and for the condensate fraction, in Section\, 5.

 \section{Evaporative cooling and time-dependent entropy}
 To achieve the phase transition to the condensate, successive evaporative cooling is used, thereby removing high-velocity atoms. In the course of the subsequent equilibration, the number of condensed particles rises due to a transfer from the nonlinear kinetic region into the coherent region \cite{svi91,kss92,kas97} and an isotropic tail \cite{an95,BookPitaevskii} develops in the ultraviolet, smearing out the sharp cut that corresponds to evaporative cooling.
 
%__________________________________________
%\section{Time-dependent entropy}
Solutions of the nonlinear boson diffusion equation fulfil the physically reasonable condition \cite{gw20} that the entropy drops in the course of an evaporative cooling step and then increases with time towards the equilibrium value that is determined by the final Bose-Einstein distribution, such that there is an interplay of cooling and re-thermalization. 

To compute the entropy, we first consider a bosonic system in statistical equilibrium that will be the limit of the time-dependent case for $t\rightarrow\infty$,
\begin{align}
\label{eq:gen_ent}
S_\text{eq} = \int g(\epsilon)\Bigl [ \bigl (1+n_\text{eq}(\epsilon)\bigr) \ln \bigl(1+n_\text{eq}(\epsilon)\bigr)\\ \nonumber
 - n_\text{eq}(\epsilon) \ln n_\text{eq}(\epsilon)\Bigr]\,\text{d}\epsilon\,.
\end{align}
Here the density of states $g(\epsilon)$ for a three-dimensional isotropic Bose gas without external potential is given by
\begin{align}
    g(\epsilon) = g_0 \, \sqrt{\epsilon}
\end{align}
with
% $(\hbar=c=k_\text{B}=1)$
\begin{align}
g_0=(2m)^{3/2}\,V/(4\pi^2)\,,
   % g_0 = \frac{V}{4 \pi^2} \left(\frac{2 m}{\hbar^2}\right)^{\frac{3}{2}}\,.
\end{align}
as obtained from the substitution of a summation over the quantum numbers of the associated states with an energy integration \cite{BookPitaevskii}. For a harmonic oscillator potential the dependence is $g(\epsilon) = g^\text{HO}_0 \epsilon^2$.

In order to check the numerical integration in Eq.\,(\ref{eq:gen_ent}) against an exact result, we derive $S_\text{eq}^{\mu\rightarrow 0}$ for a free three-dimensional Bose gas as
%and the density of states $g(\epsilon)=g_0\sqrt{\epsilon}$  as
\begin{align}
\label{eq:exact_s}
S_{\text{eq}}^{\mu\rightarrow 0}/g_0 = T^{3/2}\Biggl( \zeta\Bigl(\frac{5}{2}\Bigr) \Gamma\Bigl(\frac{5}{2}\Bigr)\qquad\qquad\qquad\\ \nonumber
+ \frac{\sqrt{\pi}}{2} \sum_{k=1}^{\infty} 
\frac{1}{k} \Bigl[ \zeta(3/2,k) -  \zeta(3/2,k+1)\Bigr]\Biggr)
\end{align}
with the Hurwitz zeta function $\zeta(s,q)$, yielding  $S_{\text{eq}}^{\mu\rightarrow 0}/(g_0\,T^{3/2}) = 2.9721553$ as compared to the result $2.97216$ of a numerical integration from zero to infinity.

If the final temperature  $T_\text{f}$ is below the initial one as is the case for cooling, the entropy drops below the initial equilibrium value and then rises again towards the final value in the course of the equilibration process. For  $T_\text{f}$ below critical value for condensate formation $T_\text{c}$,  particles also occupy the condensed state, but the total entropy still equals the entropy of the atoms in the thermal cloud. As has been emphasized in Ref.\,\cite{scul18}, this is the case even though the entropy of the particles in the ground state is nonzero, because the latter is cancelled by the so-called correlation entropy due to the fixed number of particles distributed among the quantum states. 

It is therefore sufficient for a calculation of the total time-dependent entropy $S(t)$ to consider only the thermal cloud. The analytical solutions based on Eqs.\,(\ref{eq:Nformula}) and (\ref{eq:Zarbtemp}) for constant chemical potential $\mu$ and boundary conditions at the singularity $\epsilon=\mu$  are used. The  entropy in a bosonic system
for an average number of particles $n(\epsilon,t)$ per single-particle state can be reformulated \cite{yam86,gw20} as
\begin{eqnarray}
S(t)=\int_0^\infty g(\epsilon)\Bigl[\ln\bigl(1+n(\epsilon,t)\bigr)\qquad\qquad\notag\\
+n(\epsilon,t)\ln\bigl(1+1/n(\epsilon,t)\bigr)\Bigr]\text{d}\epsilon\,.
\label{entropy}
\end{eqnarray}
Again, the spatial dimensionality and the external confinement enter the present formulation only through the density of states. The properties of the trapping potential and its effect on the density of states in evaporative cooling have been previously discussed in Ref.\,\cite{dmk95}}.
			\begin{figure}
	\centering
	\includegraphics[scale=0.86]{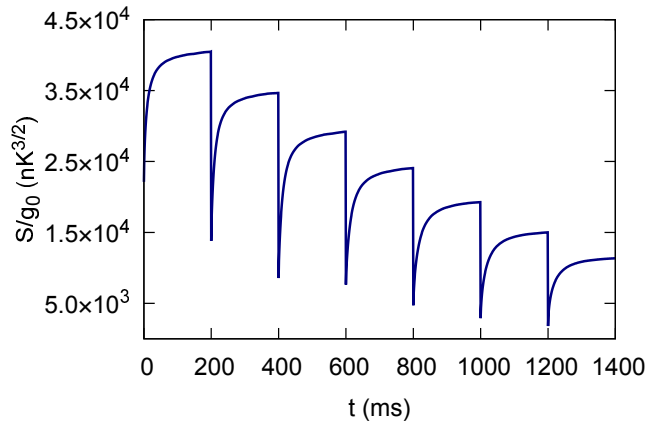}% Here is how to import EPS art
	\caption{\label{fig4}    Time evolution of the entropy $S(t)/g_0$ in an equilibrating Bose gas for $k=0-6$ evaporative cooling steps starting with $T_\text{i}=1000$ nK and $T_k = 1000 \times (0.8)^k$ nK as calculated
	from the analytical solution of the NBDE Eq.\,(\ref{bose}) with constant transport coefficients and constant chemical potential $\mu = - 8$ nK, solid curves. In each step, $S(t)$ drops instantaneously due to cooling and then rises in the course of re-thermalization. }	\end{figure}
	
				\begin{figure}
	\centering
	\includegraphics[scale=0.56]{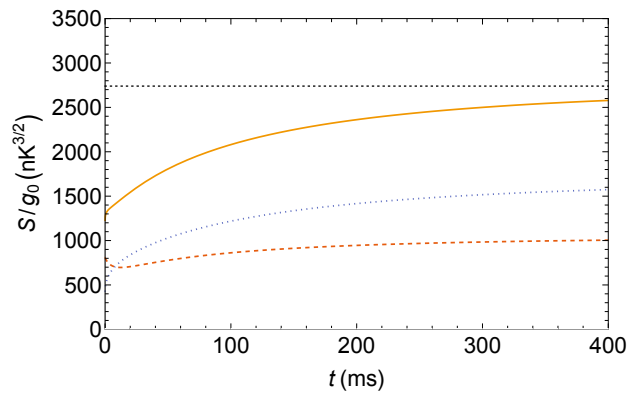}% Here is how to import EPS art
	\caption{\label{fig5}    Time evolution of the entropy $S(t)/g_0$ in an equilibrating Bose gas in the course of single-step evaporative cooling from $T_\text{i}=240$ nK with a cut at $\epsilon_\text{i}=80$\,nK to $T_\text{f}=100$ nK for $\mu=-8$\,nK as calculated
	from the analytical solution of the NBDE Eq.\,(\ref{bose}), solid curve.  {$S(t=0)$ is the entropy following evaporative cooling}. The dashed curve is the wave entropy, the dotted curve the particle entropy, and the dot-dashed curve the result from the linear relaxation ansatz \cite{gw20}. The dotted horizontal line indicates the equilibrium value  {at $T_\text{f}=100$ nK} calculated from Eq.\,(\ref{eq:gen_ent}).}
	\end{figure}
Results for successive cooling as displayed in Fig.\;\ref{fig4} show that the entropy decreases in each cooling step and subsequently rises due to re-thermalization. Correspondingly, for $T_\text{f}<T_\text{c}$ the number of condensed particles rises due to a transfer from the nonlinear kinetic region into the condensate \cite{svi91,kas97}, as will be discussed in the next section. In the example shown here
with $T_\text{i}\simeq 1000$ nK and $T_k = 1000 \times (0.8)^k$ nK, $\epsilon_{\text{cut},k} = 1000 \times (0.7)^{k+1}$ nK, $k = 0\ldots6$. 
The entropy of the initial thermal distribution -- without the cut -- is about four times larger than the one of the final distribution.
%with $T_\text{i}\simeq 240$ ((?)) nK and $T_\text{f}\simeq 100$ ((?)) nK, the entropy of the initial thermal distribution -- without the cut -- is about xxx times larger than the one of the final distribution.
% whereas the initial nonequilibrium distribution -- with the cut at $\epsilon=\epsilon_\text{i}=0.3 T_\text{i}$ ((?))\,nK -- carries only half the entropy of the final equilibrium distribution at %the lower temperature $T_\text{f}$. 

The first term  {in the entropy Eq.\,(\ref{entropy})}  is referred to as wave entropy, it yields the largest contribution when the single-particle state is occupied by many particles, as in the IR. The second term is the particle entropy, which is more relevant in case of low occupation $n(\epsilon,t)<1$, as in the UV \cite{yam86,gw20}.
%For $n=1$, both terms become equal.
%The bosonic entropy can also be expressed as
%\begin{eqnarray}
%S(t)=\int_0^\infty g(\epsilon)\Bigl[\bigl(1+n(\epsilon,t)\bigr)\,\ln\bigl(1+n(\epsilon,t)\bigr)\notag\\
%-n(\epsilon,t)\ln\bigl(n(\epsilon,t)\bigr)\Bigr]\text{d}\epsilon\,,
%\end{eqnarray}
%but here the distinction of wave- and particle entropy is not obvious.

Results for the time-dependent  contributions to the entropy are shown in Fig.\,\ref{fig5}, calculated with the analytical solutions of the NBDE. Here, the equilibrium value of the entropy at the initial temperature $T_\text{i}\simeq 240$ nK and for $\mu=-8$ \,nK is $S_\text{eq}/g_0\,(T_\text{i})\simeq10666$ (nK)$^{3/2}$. With a sharp cutoff at $\epsilon_\text{i}=80$\,nK
to account for evaporative cooling, the entropy of the initial (cooled) nonequilibrium distribution is reduced to  $S_\text{i}/g_0\,(T_\text{i})\simeq 1218$\, (nK)$^{3/2}$. 

Thermalization during the time evolution then occurs through the analytical solutions of the NBDE, and the result for the rising entropy $S(t)/g_0$ is shown in the solid curve
in Fig.\,\ref{fig5}. 
The new equilibrium value of the entropy at the final temperature $T_\text{f}\simeq 100$ nK following evaporative cooling and thermalization is $S_\text{eq}/g_0\,(T_\text{f})\simeq 2740$ \,(nK)$^{3/2}$, dotted horizontal line.
% For the given parameter set, it is reached within about 1 ms. 

The dashed curve is the wave entropy. It is most relevant for large occupation numbers, which are present at all times in the IR region $\epsilon<\epsilon_\text{i}$, and therefore, this contribution shows a rather weak time dependence. The dotted curve is the particle entropy, which is initially smaller than the wave entropy, because the occupation in the UV beyond the cut
%with $\epsilon>\epsilon_{i}$
is negligible at small times. In the course of thermalization, however, it exceeds the wave entropy at $t\simeq 10$ ms for the parameter set used in this calculation, and rises subsequently. Hence, the entropy at large times is mostly determined by the slowly rising contribution of the thermal tail.

In contrast, results for the corresponding linear relaxation-time approximation \cite{gw20} (dot-dashed curve in Fig.\,\ref{fig5}) reach the thermal limit that is given by 
Eq.\,(\ref{eq:gen_ent}) much faster because an exponential buildup of the thermal tail towards $n_\text{eq}$ is enforced.
The entropy approaches the same equilibrium value, but at shorter times than the nonlinear solution.

It is noted that the final state does not change if both transport coefficients are scaled by the same amount because $T_\text{f}=-D/v$, but the time scale varies since $\tau_\text{eq}=4D/(9v^2)$. By measuring the time it takes for a given system to thermalize we can thus determine the absolute magnitude of the transport coefficients. To obtain the time scale from first principles, a microscopic calculation of the transport coefficients from a many-body theory is required, which is beyond the scope of our present approach.

Up to now, the results for the time-dependent occupation-number distributions and the entropy have been discussed for constant chemical potential. When computing the condensate fraction, however, particle-number conservation must be considered and $\mu$ becomes time-dependent, see the following section.   	
%It will be interesting to investigate the time dependence of the entropy in a nonlinear model that is not restricted to constant transport coefficients, although 
%With a time dependence in the diffusion coefficient, in particular, the nonlinear diffusion does not set in with the full strength at short times, and the rise of the entropy with time is %expected to be even slower. 
 %it is unlikely that exact solutions of the NBDE with variable transport coefficients can be found.

 \begin{figure}
	\centering
	\includegraphics[scale=0.9]{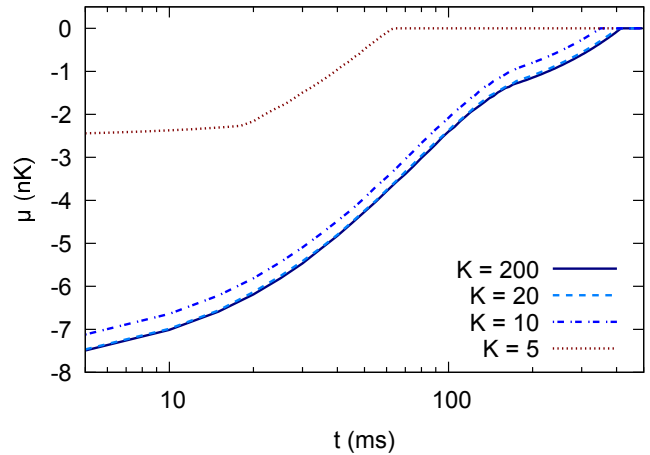}% Here is how to import EPS art
	\caption{\label{fig6}  Chemical potential $\mu(t)$ in the course of evaporative cooling from $T_\text{i}=240$ nK to $T_\text{f}=100$ nK as calculated
	based on particle-number conservation from the analytical solution of the NBDE Eq.\,(\ref{bose}) with $k_\text{max}\equiv K= 5, 10, 20, 200$ in the series expansion of the exact solution with $\epsilon_\text{i}=100$\,nK and $\mu_\text{i}=-8$\,nK.} 
	%\label{fig6} 
	\end{figure}
\section{Time-dependent condensate fraction}
{We proceed to calculate the rate of increase of atoms in the condensate $N_\text{c}(t)$ for conserved total particle number $N = N_\text{c}(t)+N_\text{th}(t)$ and $T_\text{f}<T_\text{c}$
based on the exact analytic NBDE-solutions Eq.\,(\ref{eq:Nformula}) for any given $\mu$. Particle-number conservation -- which is a necessary condition for condensate formation to occur -- requires a time-dependent chemical potential, with $\mu(t)\rightarrow0$ corresponding to the condensed state. 
 %We consider this further in Section\,V, after having computed the time-dependent entropy in Section\,IV.

 To compute $\mu(t)$ in accordance with overall particle-number conservation, we
 note that the particle number of the thermal cloud is given by
\EQ{eq:n_thermal}
%{N_\text{th}(t) = \int_0^{\infty} g(\epsilon)\, n(\epsilon-\mu(t),t)\, \text{d}\epsilon
{N_\text{th}(t) = \int_0^{\infty} g(\epsilon)\, \tilde{n}(\epsilon,t)\, \text{d}\epsilon
}{,}
where $\tilde{n}(\epsilon,t)$ is the distribution function with the time-dependent value of $\mu\equiv \mu(t)$. For each time step,
$\mu(t)$ is calculated such that the particle number is conserved,  $N = N_\text{th}$, until
$\mu$ reaches the value of zero and the largest possible value of $N_\text{th}$ is attained. This instant marks
the moment of condensation and any future difference between the initial particle number and the thermal cloud particle number
is accounted for by the condensate 
\EQ{}
{N - N_\text{th}(t) \equiv N_c(t)\,, \quad \text{for } \mu=0
}{.}
A typical time evolution of $\mu(t)$ is displayed in Fig.\,\ref{fig6}, for different values of $k_\text{max}\equiv K$ to show the convergence. Here the initial chemical potential is
$\mu_\text{i}=-8$ nK, the final one  $\mu_\text{f}=0$. With increasing $k_\text{max}$ the accuracy increases rapidly,  precise results are obtained already for $k_\text{max}=20$. For the parameters used in Fig.\,\ref{fig6}, condensation sets in at around $400$ ms. 
As is obvious from Eq.\,\eqref{eq:n_thermal}, $\mu(t)$ and hence, the onset of condensation, also depends on the density of states.

Due to the statistical properties of the bosonic system that are encoded in the NBDE, no condensate forms if the final temperature remains above the critical value. We have confirmed this in calculations for different final temperatures $T_\text{f}$, with the condensate fraction $N_\text{c}(t)/N=0\, \forall \, t \land T_\text{f}>T_\text{c}$.
It is also self-evident because $T_\text{c}$ is derived from the Bose-Einstein equilibrium solution, which is the stationary limit of the NBDE for $t\rightarrow\infty$.
					\begin{figure}
	\centering
	\includegraphics[scale=0.88]{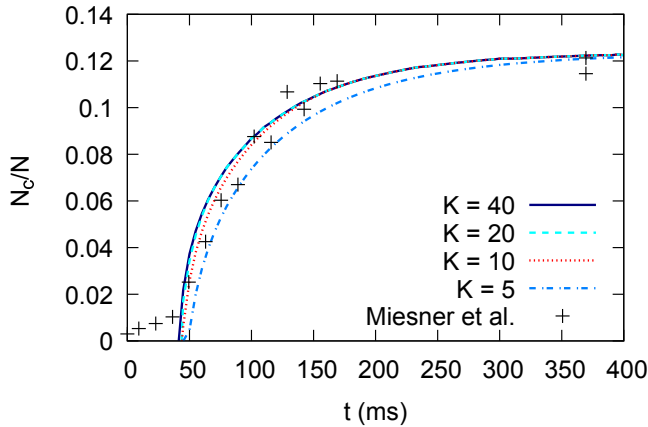}% Here is how to import EPS art
	\caption{\label{fig7}   Condensate fraction $N_\text{c}(t)/N$ in an equilibrating Bose gas of $^{23}$Na subsequent to fast evaporative cooling in a single step from $T_\text{i}=876$ nK to $T_\text{f}=750$ nK as calculated
	from the analytical solution of the NBDE Eq.\,(\ref{bose}) with $k_\text{max}\equiv K=5, 10, 20, 40$ in the series expansion of the exact solution, cutoff energy 
	$\epsilon_\text{i}=2190$\,nK, $\mu_\text{i}=-8$\,nK, and the density of states for a free Bose gas.
    %The dotted horizontal line indicates the equilibrium value  {at $T_\text{f}=yyy$ nK} calculated with the density of states in a harmonic trap. 
    The transport coefficients are $D = 3750$\,(nK)$^2$\,ms$^{-1},~\,v = -5$\,nK$\,$ms$^{-1}$.
    The MIT data for the condensate fraction (crosses, no error bars) are from Ref.\,\cite{miesner_bosonic_1998}.}
	\end{figure}   
    					\begin{figure}
	\centering
	\includegraphics[scale=0.88]{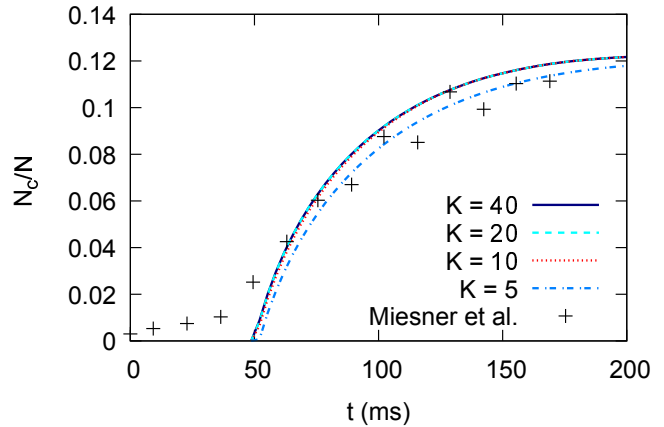}% Here is how to import EPS art
	\caption{\label{fig8}   Condensate fraction $N_\text{c}(t)/N$ as in Fig.\,\ref{fig7}, following evaporative cooling (RF-sweep) in a single step from $T_\text{i}=876$ nK to $T_\text{f}=660$ nK as calculated
	from the analytical solution of the NBDE Eq.\,(\ref{bose}) with $k_\text{max}\equiv K=5, 10, 20, 40$, $\epsilon_\text{i}=1810$\,nK, $\mu_\text{i}=-18$\,nK, and the density of states in a harmonic trap.
Due to the constant transport coefficients, the diffusion into the condensate starts instantaneously. 
The transport coefficients are $D = 3300$\,(nK)$^2$\,ms$^{-1},~\,v = -5$\,nK$\,$ms$^{-1}$.
The MIT data for $^{23}$Na are from Ref.\,\cite{miesner_bosonic_1998}, with an enlarged time scale. }
	\end{figure}
In order to compare our results for $T_\text{f}<T_\text{c}$ to data we use the MIT measurements from Ref.\,\cite{miesner_bosonic_1998} with parameters from
Ref.\,\cite{bzs00}, where a nonlinear Schr\"odinger equation coupled to a 
Boltzmann-like quantum collision term has been solved numerically. 

In particular, the final number of 
condensate atoms 
%and the time evolution of the condensate fraction 
is taken from Ref.\,\cite{miesner_bosonic_1998}. The initial 
number of atoms in the trap is unknown but in Ref.\,\cite{bzs00}, $N_0$ = \num{60e6} has been used
in order to achieve agreement of a numerical simulation and the data. We adopt their values of $T_\text{i} = 876$\,{nK} and $\epsilon_\text{i}/T_\text{i}$ = \num{2.5}. In contrast to Ref.\,\cite{bzs00} our treatment also requires the specification of the final 
temperature which we infer from energy conservation in a numerical calculation, yielding $T_\text{f} =  750$\,nK for the free gas, and  $T_\text{f} =  660$\,nK with the density of states in a harmonic oscillator potential as in the MIT trap.

After solving the NBDE for $n(\epsilon,t)$ at fixed initial chemical potential we calculate $\mu(t)$ at all times and finally
compute $N_c(t)/N_\text{tot}$ which is shown in Fig.\,\ref{fig7} together with the convergence of the analytic solution for different values of $k_\text{max}$ in the
series expansion of the exact analytic solution. In this result we have used the density of states for a free gas, to be compared with the one for a harmonic potential
 in Fig.\,\ref{fig8} (here the timescale is enlarged to emphasize the small-time region).

In both cases, the time axis has been shifted to give the best possible fit. The transport coefficients were adapted
such that the observed time scale was reproduced. With these adjustments, data and model result are found to agree.
The solutions are steeper for the free gas at small times compared to the harmonic-oscillator density of states.

One feature both solutions fail to account for, however,
is the measured initial slow increase of the condensate fraction in the first {30}\,{ms}, which cannot be reproduced in our model with constant
transport coefficients. Should the small-time gradual rise of the condensate persist in future measurements, we can adapt the model using
time-dependent transport coefficients, such that the diffusion into the condensate builds up slowly with time: For constant drift and diffusion coefficients,
the diffusion into the condensate sets in instantaneously with full strength and without a seed condensate.

It is recognized that the experiments deal with an inhomogeneous trapped Bose gas, whereas the theory in the present form is homogeneous. In particular, we consider the difference between a finite system with and without a trapping potential only through the difference in the densities of states. This is certainly a limitation of the model, it is not meant to present a full description of the dynamics of condensate formation. But it yields physically reasonable results on a purely nonequilibrium-statistical basis.

\section{Conclusion}
Exact solutions of the nonlinear boson diffusion equation which take account of the singularity in the initial conditions at $\epsilon=\mu<0$ and the necessary boundary conditions at the singularity have been explored in this work. The analytical solutions for constant transport coefficients are compared in detail with numerical solutions of the NBDE obtained with Matlab and agreement is found. The exact solutions are used
to calculate the time-dependent entropy in a cooling Bose gas. Whereas cooling drastically reduces the entropy, the subsequent re-thermalization causes a gradual increase towards the equilibrium value, which coincides with the Bose-Einstein result and is significantly below the initial value before cooling. Wave and particle entropy are discussed. Together with particle-number conservation and a time-dependent chemical potential, the condensate fraction in a trap has been calculated as a function of time.

The solutions of the NBDE are thus shown to properly describe equilibration processes that occur in quantum gases in the course of evaporative cooling, and the associated condensate formation for $T_\text{f}<T_\text{c}$.
They are specifically applied to  evaporative cooling and condensate formation in $^{23}$Na. Agreement with MIT data for the time-dependent condensate fraction is found.
The convergence properties of the infinite series expansion in the analytic solution of the NBDE have been tested.

In this work, we have not aimed at a detailed treatment of the quantum-mechanical properties of the condensate, which is coupled to the time-dependent evolution of the thermal cloud. Instead, the characteristic nonequilibrium-statistical features of the bosonic system and in particular, the boson stimulation that is encoded in the NBDE, have been used together with particle-number conservation to infer the time-dependent evolution of the condensate fraction.

Further refinements of the model such as variable transport coefficients to describe a gradual buildup of the condensate, and the coupling to the quantum fluctuations or excitations of the condensate are conceivable, but are unlikely to allow for analytic solutions. Extensions of the NBDE itself to higher dimensions in order to better account for anisotropic systems should also be tested. 

More extensive comparisons of the results to new data from cold-atom experiments with an improved signal-to-noise ratio and systematic as well as statistical error bars would be most welcome. In particular, time-dependent measurements for other bosonic alkali atoms such as rubidium or lithium with substantially different timescales for condensate formation and equilibration are very desirable. This could, in particular, offer the opportunity to learn more about the initial BEC formation phase in the first 50 ms. Here, the quantum Boltzmann-type approaches often use an artificial seed condensate. Although this is needed to trigger condensate growth and achieve agreement with the data, it seems difficult to develop a sound physical justification for the initial condensate occupation. In contrast, in the NBDE approach no seed is needed. \\
%\newpage
%\onecolumngrid
\acknowledgments
%\noindent
%\bf{Acknowledgements}
%\rm

%\noindent
We thank both referees for questions and comments, and Joseph Indekeu of Leuven University for the editorial work.
%We are grateful to Andreas Mielke for discussions and remarks.
%\twocolumngrid
%\newpage
\bibliography{gw_20}

%apsrev4-2.bst 2019-01-14 (MD) hand-edited version of apsrev4-1.bst
%Control: key (0)
%Control: author (8) initials jnrlst
%Control: editor formatted (1) identically to author
%Control: production of article title (0) allowed
%Control: page (0) single
%Control: year (1) truncated
%Control: production of eprint (0) enabled
\begin{thebibliography}{32}%
\makeatletter
\providecommand \@ifxundefined [1]{%
 \@ifx{#1\undefined}
}%
\providecommand \@ifnum [1]{%
 \ifnum #1\expandafter \@firstoftwo
 \else \expandafter \@secondoftwo
 \fi
}%
\providecommand \@ifx [1]{%
 \ifx #1\expandafter \@firstoftwo
 \else \expandafter \@secondoftwo
 \fi
}%
\providecommand \natexlab [1]{#1}%
\providecommand \enquote  [1]{``#1''}%
\providecommand \bibnamefont  [1]{#1}%
\providecommand \bibfnamefont [1]{#1}%
\providecommand \citenamefont [1]{#1}%
\providecommand \href@noop [0]{\@secondoftwo}%
\providecommand \href [0]{\begingroup \@sanitize@url \@href}%
\providecommand \@href[1]{\@@startlink{#1}\@@href}%
\providecommand \@@href[1]{\endgroup#1\@@endlink}%
\providecommand \@sanitize@url [0]{\catcode `\\12\catcode `\$12\catcode
  `\&12\catcode `\#12\catcode `\^12\catcode `\_12\catcode `\%12\relax}%
\providecommand \@@startlink[1]{}%
\providecommand \@@endlink[0]{}%
\providecommand \url  [0]{\begingroup\@sanitize@url \@url }%
\providecommand \@url [1]{\endgroup\@href {#1}{\urlprefix }}%
\providecommand \urlprefix  [0]{URL }%
\providecommand \Eprint [0]{\href }%
\providecommand \doibase [0]{https://doi.org/}%
\providecommand \selectlanguage [0]{\@gobble}%
\providecommand \bibinfo  [0]{\@secondoftwo}%
\providecommand \bibfield  [0]{\@secondoftwo}%
\providecommand \translation [1]{[#1]}%
\providecommand \BibitemOpen [0]{}%
\providecommand \bibitemStop [0]{}%
\providecommand \bibitemNoStop [0]{.\EOS\space}%
\providecommand \EOS [0]{\spacefactor3000\relax}%
\providecommand \BibitemShut  [1]{\csname bibitem#1\endcsname}%
\let\auto@bib@innerbib\@empty
%</preamble>
\bibitem [{\citenamefont {Anderson}\ \emph {et~al.}(1995)\citenamefont
  {Anderson}, \citenamefont {Ensher}, \citenamefont {Matthews}, \citenamefont
  {Wieman},\ and\ \citenamefont {Cornell}}]{an95}%
  \BibitemOpen
  \bibfield  {author} {\bibinfo {author} {\bibfnamefont {M.~H.}\ \bibnamefont
  {Anderson}}, \bibinfo {author} {\bibfnamefont {J.~R.}\ \bibnamefont
  {Ensher}}, \bibinfo {author} {\bibfnamefont {M.~R.}\ \bibnamefont
  {Matthews}}, \bibinfo {author} {\bibfnamefont {C.~E.}\ \bibnamefont
  {Wieman}},\ and\ \bibinfo {author} {\bibfnamefont {E.~A.}\ \bibnamefont
  {Cornell}},\ }\bibfield  {title} {\bibinfo {title} {Observation of
  {Bose-Einstein} condensation in a dilute atomic vapor},\ }\href@noop {}
  {\bibfield  {journal} {\bibinfo  {journal} {Science}\ }\textbf {\bibinfo
  {volume} {269}},\ \bibinfo {pages} {198} (\bibinfo {year}
  {1995})}\BibitemShut {NoStop}%
\bibitem [{\citenamefont {Davis}\ \emph
  {et~al.}(1995{\natexlab{a}})\citenamefont {Davis}, \citenamefont {Mewes},
  \citenamefont {Andrews}, \citenamefont {van Druten}, \citenamefont {Durfee},
  \citenamefont {Kurn},\ and\ \citenamefont {Ketterle}}]{ket95}%
  \BibitemOpen
  \bibfield  {author} {\bibinfo {author} {\bibfnamefont {K.~B.}\ \bibnamefont
  {Davis}}, \bibinfo {author} {\bibfnamefont {M.-O.}\ \bibnamefont {Mewes}},
  \bibinfo {author} {\bibfnamefont {M.~R.}\ \bibnamefont {Andrews}}, \bibinfo
  {author} {\bibfnamefont {N.~J.}\ \bibnamefont {van Druten}}, \bibinfo
  {author} {\bibfnamefont {D.~S.}\ \bibnamefont {Durfee}}, \bibinfo {author}
  {\bibfnamefont {D.~M.}\ \bibnamefont {Kurn}},\ and\ \bibinfo {author}
  {\bibfnamefont {W.}~\bibnamefont {Ketterle}},\ }\bibfield  {title} {\bibinfo
  {title} {{Bose-Einstein} condensation in a gas of sodium atoms},\ }\href@noop
  {} {\bibfield  {journal} {\bibinfo  {journal} {Phys. Rev. Lett.}\ }\textbf
  {\bibinfo {volume} {75}},\ \bibinfo {pages} {3969} (\bibinfo {year}
  {1995}{\natexlab{a}})}\BibitemShut {NoStop}%
\bibitem [{\citenamefont {Bradley}\ \emph {et~al.}(1995)\citenamefont
  {Bradley}, \citenamefont {Sackett}, \citenamefont {Tollett},\ and\
  \citenamefont {Hulet}}]{hul95}%
  \BibitemOpen
  \bibfield  {author} {\bibinfo {author} {\bibfnamefont {C.~C.}\ \bibnamefont
  {Bradley}}, \bibinfo {author} {\bibfnamefont {C.~A.}\ \bibnamefont
  {Sackett}}, \bibinfo {author} {\bibfnamefont {J.~J.}\ \bibnamefont
  {Tollett}},\ and\ \bibinfo {author} {\bibfnamefont {R.~G.}\ \bibnamefont
  {Hulet}},\ }\bibfield  {title} {\bibinfo {title} {Evidence of {Bose-Einstein}
  condensation in an atomic gas with attractive interactions},\ }\href@noop {}
  {\bibfield  {journal} {\bibinfo  {journal} {Phys. Rev. Lett.}\ }\textbf
  {\bibinfo {volume} {75}},\ \bibinfo {pages} {1687} (\bibinfo {year}
  {1995})}\BibitemShut {NoStop}%
\bibitem [{\citenamefont {Bradley}\ \emph {et~al.}(1997)\citenamefont
  {Bradley}, \citenamefont {Sackett},\ and\ \citenamefont {Hulet}}]{hul97}%
  \BibitemOpen
  \bibfield  {author} {\bibinfo {author} {\bibfnamefont {C.~C.}\ \bibnamefont
  {Bradley}}, \bibinfo {author} {\bibfnamefont {C.~A.}\ \bibnamefont
  {Sackett}},\ and\ \bibinfo {author} {\bibfnamefont {R.~G.}\ \bibnamefont
  {Hulet}},\ }\bibfield  {title} {\bibinfo {title} {{Bose-Einstein}
  condensation of lithium: Observation of limited condensate number},\
  }\href@noop {} {\bibfield  {journal} {\bibinfo  {journal} {Phys. Rev. Lett.}\
  }\textbf {\bibinfo {volume} {78}},\ \bibinfo {pages} {985} (\bibinfo {year}
  {1997})}\BibitemShut {NoStop}%
\bibitem [{\citenamefont {Gardiner}\ \emph {et~al.}(1997)\citenamefont
  {Gardiner}, \citenamefont {Zoller}, \citenamefont {Ballagh},\ and\
  \citenamefont {Davis}}]{gz97}%
  \BibitemOpen
  \bibfield  {author} {\bibinfo {author} {\bibfnamefont {C.~W.}\ \bibnamefont
  {Gardiner}}, \bibinfo {author} {\bibfnamefont {P.}~\bibnamefont {Zoller}},
  \bibinfo {author} {\bibfnamefont {R.~J.}\ \bibnamefont {Ballagh}},\ and\
  \bibinfo {author} {\bibfnamefont {M.~J.}\ \bibnamefont {Davis}},\ }\bibfield
  {title} {\bibinfo {title} {Kinetics of {Bose-Einstein} condensation in a
  trap},\ }\href@noop {} {\bibfield  {journal} {\bibinfo  {journal} {Phys. Rev.
  Lett}\ }\textbf {\bibinfo {volume} {79}},\ \bibinfo {pages} {1793} (\bibinfo
  {year} {1997})}\BibitemShut {NoStop}%
\bibitem [{\citenamefont {Gardiner}\ and\ \citenamefont
  {Zoller}(1997)}]{gardiner_quantum_1997}%
  \BibitemOpen
  \bibfield  {author} {\bibinfo {author} {\bibfnamefont {C.~W.}\ \bibnamefont
  {Gardiner}}\ and\ \bibinfo {author} {\bibfnamefont {P.}~\bibnamefont
  {Zoller}},\ }\bibfield  {title} {\bibinfo {title} {Quantum kinetic theory:
  {{A}} quantum kinetic master equation for condensation of a weakly
  interacting {{Bose}} gas without a trapping potential},\ }\href@noop {}
  {\bibfield  {journal} {\bibinfo  {journal} {Phys. Rev. A}\ }\textbf {\bibinfo
  {volume} {55}},\ \bibinfo {pages} {2902} (\bibinfo {year}
  {1997})}\BibitemShut {NoStop}%
\bibitem [{\citenamefont {Bijlsma}\ \emph {et~al.}(2000)\citenamefont
  {Bijlsma}, \citenamefont {Zaremba},\ and\ \citenamefont {Stoof}}]{bzs00}%
  \BibitemOpen
  \bibfield  {author} {\bibinfo {author} {\bibfnamefont {M.~J.}\ \bibnamefont
  {Bijlsma}}, \bibinfo {author} {\bibfnamefont {E.}~\bibnamefont {Zaremba}},\
  and\ \bibinfo {author} {\bibfnamefont {H.~T.~C.}\ \bibnamefont {Stoof}},\
  }\bibfield  {title} {\bibinfo {title} {Condensate growth in trapped {Bose}
  gases},\ }\href@noop {} {\bibfield  {journal} {\bibinfo  {journal} {Phys.
  Rev. A}\ }\textbf {\bibinfo {volume} {62}},\ \bibinfo {pages} {063609}
  (\bibinfo {year} {2000})}\BibitemShut {NoStop}%
\bibitem [{\citenamefont {Miesner}\ \emph {et~al.}(1998)\citenamefont
  {Miesner}, \citenamefont {Stamper-Kurn}, \citenamefont {Andrews},
  \citenamefont {Durfee}, \citenamefont {Inouye},\ and\ \citenamefont
  {Ketterle}}]{miesner_bosonic_1998}%
  \BibitemOpen
  \bibfield  {author} {\bibinfo {author} {\bibfnamefont {H.-J.}\ \bibnamefont
  {Miesner}}, \bibinfo {author} {\bibfnamefont {D.~M.}\ \bibnamefont
  {Stamper-Kurn}}, \bibinfo {author} {\bibfnamefont {M.~R.}\ \bibnamefont
  {Andrews}}, \bibinfo {author} {\bibfnamefont {D.~S.}\ \bibnamefont {Durfee}},
  \bibinfo {author} {\bibfnamefont {S.}~\bibnamefont {Inouye}},\ and\ \bibinfo
  {author} {\bibfnamefont {W.}~\bibnamefont {Ketterle}},\ }\bibfield  {title}
  {\bibinfo {title} {Bosonic stimulation in the formation of a
  {Bose}-{Einstein} condensate},\ }\href@noop {} {\bibfield  {journal}
  {\bibinfo  {journal} {Science}\ }\textbf {\bibinfo {volume} {279}},\ \bibinfo
  {pages} {1005} (\bibinfo {year} {1998})}\BibitemShut {NoStop}%
\bibitem [{\citenamefont {Eckern}(1984)}]{eckern84}%
  \BibitemOpen
  \bibfield  {author} {\bibinfo {author} {\bibfnamefont {U.}~\bibnamefont
  {Eckern}},\ }\bibfield  {title} {\bibinfo {title} {Relaxation processes in a
  condensed {Bose} gas},\ }\href@noop {} {\bibfield  {journal} {\bibinfo
  {journal} {J. Low Temp. Phys.}\ }\textbf {\bibinfo {volume} {54}},\ \bibinfo
  {pages} {333} (\bibinfo {year} {1984})}\BibitemShut {NoStop}%
\bibitem [{\citenamefont {Kirkpatrick}\ and\ \citenamefont
  {Dorfman}(1985)}]{kd85}%
  \BibitemOpen
  \bibfield  {author} {\bibinfo {author} {\bibfnamefont {T.~R.}\ \bibnamefont
  {Kirkpatrick}}\ and\ \bibinfo {author} {\bibfnamefont {J.~R.}\ \bibnamefont
  {Dorfman}},\ }\bibfield  {title} {\bibinfo {title} {Transport coefficients in
  a dilute but condensed {Bose} gas},\ }\href@noop {} {\bibfield  {journal}
  {\bibinfo  {journal} {J. Low Temp. Phys.}\ }\textbf {\bibinfo {volume}
  {58}},\ \bibinfo {pages} {301} (\bibinfo {year} {1985})}\BibitemShut
  {NoStop}%
\bibitem [{\citenamefont {Stoof}(1997)}]{sto97}%
  \BibitemOpen
  \bibfield  {author} {\bibinfo {author} {\bibfnamefont {H.~T.~C.}\
  \bibnamefont {Stoof}},\ }\bibfield  {title} {\bibinfo {title} {Initial stages
  of {Bose-Einstein} condensation},\ }\href@noop {} {\bibfield  {journal}
  {\bibinfo  {journal} {Phys. Rev. Lett.}\ }\textbf {\bibinfo {volume} {78}},\
  \bibinfo {pages} {768} (\bibinfo {year} {1997})}\BibitemShut {NoStop}%
\bibitem [{\citenamefont {Stoof}(1999)}]{stoof_coherent_1999}%
  \BibitemOpen
  \bibfield  {author} {\bibinfo {author} {\bibfnamefont {H.~T.~C.}\
  \bibnamefont {Stoof}},\ }\bibfield  {title} {\bibinfo {title} {Coherent
  versus incoherent dynamics during {Bose-Einstein} condensation in atomic
  gases},\ }\href@noop {} {\bibfield  {journal} {\bibinfo  {journal} {J. Low
  Temp. Phys.}\ }\textbf {\bibinfo {volume} {114}},\ \bibinfo {pages} {11}
  (\bibinfo {year} {1999})}\BibitemShut {NoStop}%
\bibitem [{\citenamefont {Zaremba}\ \emph {et~al.}(1999)\citenamefont
  {Zaremba}, \citenamefont {Nikuni}, \citenamefont {Griffin},\ and\
  \citenamefont {Low}}]{zng99}%
  \BibitemOpen
  \bibfield  {author} {\bibinfo {author} {\bibfnamefont {E.}~\bibnamefont
  {Zaremba}}, \bibinfo {author} {\bibfnamefont {T.}~\bibnamefont {Nikuni}},
  \bibinfo {author} {\bibfnamefont {A.}~\bibnamefont {Griffin}},\ and\ \bibinfo
  {author} {\bibfnamefont {J.}~\bibnamefont {Low}},\ }\bibfield  {title}
  {\bibinfo {title} {Dynamics of trapped {Bose} gases at finite temperatures},\
  }\href@noop {} {\bibfield  {journal} {\bibinfo  {journal} {J. Low Temp.
  Phys.}\ }\textbf {\bibinfo {volume} {116}},\ \bibinfo {pages} {277} (\bibinfo
  {year} {1999})}\BibitemShut {NoStop}%
\bibitem [{\citenamefont {Snoke}\ and\ \citenamefont {Wolfe}(1989)}]{snowo89}%
  \BibitemOpen
  \bibfield  {author} {\bibinfo {author} {\bibfnamefont {D.~W.}\ \bibnamefont
  {Snoke}}\ and\ \bibinfo {author} {\bibfnamefont {J.~P.}\ \bibnamefont
  {Wolfe}},\ }\bibfield  {title} {\bibinfo {title} {Population dynamics of a
  {Bose} gas near saturation},\ }\href@noop {} {\bibfield  {journal} {\bibinfo
  {journal} {Phys. Rev. B}\ }\textbf {\bibinfo {volume} {39}},\ \bibinfo
  {pages} {4030} (\bibinfo {year} {1989})}\BibitemShut {NoStop}%
\bibitem [{\citenamefont {Kagan}\ \emph {et~al.}(1992)\citenamefont {Kagan},
  \citenamefont {Svistunov},\ and\ \citenamefont {Shlyapnikov}}]{kss92}%
  \BibitemOpen
  \bibfield  {author} {\bibinfo {author} {\bibfnamefont {Y.~M.}\ \bibnamefont
  {Kagan}}, \bibinfo {author} {\bibfnamefont {B.~V.}\ \bibnamefont
  {Svistunov}},\ and\ \bibinfo {author} {\bibfnamefont {G.~V.}\ \bibnamefont
  {Shlyapnikov}},\ }\bibfield  {title} {\bibinfo {title} {Kinetics of {Bose}
  condensation in an interacting {Bose} gas},\ }\href@noop {} {\bibfield
  {journal} {\bibinfo  {journal} {Sov. Phys. JETP}\ }\textbf {\bibinfo {volume}
  {74}},\ \bibinfo {pages} {279} (\bibinfo {year} {1992})}\BibitemShut
  {NoStop}%
\bibitem [{\citenamefont {Semikoz}\ and\ \citenamefont
  {Tkachev}(1995)}]{setk95}%
  \BibitemOpen
  \bibfield  {author} {\bibinfo {author} {\bibfnamefont {D.~V.}\ \bibnamefont
  {Semikoz}}\ and\ \bibinfo {author} {\bibfnamefont {I.~I.}\ \bibnamefont
  {Tkachev}},\ }\bibfield  {title} {\bibinfo {title} {Kinetics of {Bose}
  condensation},\ }\href@noop {} {\bibfield  {journal} {\bibinfo  {journal}
  {Phys. Rev. Lett.}\ }\textbf {\bibinfo {volume} {74}},\ \bibinfo {pages}
  {3093} (\bibinfo {year} {1995})}\BibitemShut {NoStop}%
\bibitem [{\citenamefont {Luiten}\ \emph {et~al.}(1996)\citenamefont {Luiten},
  \citenamefont {Reynolds},\ and\ \citenamefont {Walraven}}]{lrw96}%
  \BibitemOpen
  \bibfield  {author} {\bibinfo {author} {\bibfnamefont {O.~J.}\ \bibnamefont
  {Luiten}}, \bibinfo {author} {\bibfnamefont {M.~W.}\ \bibnamefont
  {Reynolds}},\ and\ \bibinfo {author} {\bibfnamefont {J.~T.~M.}\ \bibnamefont
  {Walraven}},\ }\bibfield  {title} {\bibinfo {title} {Kinetic theory of the
  evaporative cooling of a trapped gas},\ }\href@noop {} {\bibfield  {journal}
  {\bibinfo  {journal} {Phys. Rev. A}\ }\textbf {\bibinfo {volume} {53}},\
  \bibinfo {pages} {381} (\bibinfo {year} {1996})}\BibitemShut {NoStop}%
\bibitem [{\citenamefont {Holland}\ \emph {et~al.}(1997)\citenamefont
  {Holland}, \citenamefont {Williams},\ and\ \citenamefont {Cooper}}]{hwc97}%
  \BibitemOpen
  \bibfield  {author} {\bibinfo {author} {\bibfnamefont {M.}~\bibnamefont
  {Holland}}, \bibinfo {author} {\bibfnamefont {J.}~\bibnamefont {Williams}},\
  and\ \bibinfo {author} {\bibfnamefont {J.}~\bibnamefont {Cooper}},\
  }\bibfield  {title} {\bibinfo {title} {{Bose-Einstein} condensation: Kinetic
  evolution obtained from simulated trajectories},\ }\href@noop {} {\bibfield
  {journal} {\bibinfo  {journal} {Phys. Rev. A}\ }\textbf {\bibinfo {volume}
  {55}},\ \bibinfo {pages} {3670} (\bibinfo {year} {1997})}\BibitemShut
  {NoStop}%
\bibitem [{\citenamefont {Jaksch}\ \emph {et~al.}(1997)\citenamefont {Jaksch},
  \citenamefont {Gardiner},\ and\ \citenamefont {Zoller}}]{jgz97}%
  \BibitemOpen
  \bibfield  {author} {\bibinfo {author} {\bibfnamefont {D.}~\bibnamefont
  {Jaksch}}, \bibinfo {author} {\bibfnamefont {C.~W.}\ \bibnamefont
  {Gardiner}},\ and\ \bibinfo {author} {\bibfnamefont {P.}~\bibnamefont
  {Zoller}},\ }\bibfield  {title} {\bibinfo {title} {Quantum kinetic theory.
  {II. Simulation} of the quantum {Boltzmann} master equation},\ }\href@noop {}
  {\bibfield  {journal} {\bibinfo  {journal} {Phys. Rev. A}\ }\textbf {\bibinfo
  {volume} {56}},\ \bibinfo {pages} {575} (\bibinfo {year} {1997})}\BibitemShut
  {NoStop}%
\bibitem [{\citenamefont {Wolschin}(2018{\natexlab{a}})}]{gw18}%
  \BibitemOpen
  \bibfield  {author} {\bibinfo {author} {\bibfnamefont {G.}~\bibnamefont
  {Wolschin}},\ }\bibfield  {title} {\bibinfo {title} {Equilibration in finite
  {Bose} systems},\ }\href@noop {} {\bibfield  {journal} {\bibinfo  {journal}
  {Physica A}\ }\textbf {\bibinfo {volume} {499}},\ \bibinfo {pages} {1}
  (\bibinfo {year} {2018}{\natexlab{a}})}\BibitemShut {NoStop}%
\bibitem [{\citenamefont {Wolschin}(2018{\natexlab{b}})}]{gw18a}%
  \BibitemOpen
  \bibfield  {author} {\bibinfo {author} {\bibfnamefont {G.}~\bibnamefont
  {Wolschin}},\ }\bibfield  {title} {\bibinfo {title} {An exactly solvable
  model for equilibration in bosonic systems},\ }\href@noop {} {\bibfield
  {journal} {\bibinfo  {journal} {EPL}\ }\textbf {\bibinfo {volume} {123}},\
  \bibinfo {pages} {20009} (\bibinfo {year} {2018}{\natexlab{b}})}\BibitemShut
  {NoStop}%
\bibitem [{\citenamefont {Wolschin}(2020)}]{gw20}%
  \BibitemOpen
  \bibfield  {author} {\bibinfo {author} {\bibfnamefont {G.}~\bibnamefont
  {Wolschin}},\ }\bibfield  {title} {\bibinfo {title} {Time-dependent entropy
  of a cooling {{Bose}} gas},\ }\href@noop {} {\bibfield  {journal} {\bibinfo
  {journal} {EPL}\ }\textbf {\bibinfo {volume} {129}},\ \bibinfo {pages}
  {40006} (\bibinfo {year} {2020})}\BibitemShut {NoStop}%
\bibitem [{\citenamefont {Nordheim}(1928)}]{no28}%
  \BibitemOpen
  \bibfield  {author} {\bibinfo {author} {\bibfnamefont {L.~W.}\ \bibnamefont
  {Nordheim}},\ }\bibfield  {title} {\bibinfo {title} {On the kinetic method in
  the new statistics and application in the electron theory of conductivity},\
  }\href@noop {} {\bibfield  {journal} {\bibinfo  {journal} {Proc. R. Soc.
  Lond. A}\ }\textbf {\bibinfo {volume} {119}},\ \bibinfo {pages} {689}
  (\bibinfo {year} {1928})}\BibitemShut {NoStop}%
\bibitem [{Note1()}]{Note1}%
  \BibitemOpen
  \bibinfo {note} {The derivative-term of the diffusion coefficient has been
  modified as compared to Refs.\protect \tmspace +\thinmuskip {.1667em}\cite
  {gw18,gw18a,gw20} in order to secure the correct stationary
  solution}\BibitemShut {NoStop}%
\bibitem [{\citenamefont {Rasch}\ and\ \citenamefont {Wolschin}(2020)}]{rgw20}%
  \BibitemOpen
  \bibfield  {author} {\bibinfo {author} {\bibfnamefont {N.}~\bibnamefont
  {Rasch}}\ and\ \bibinfo {author} {\bibfnamefont {G.}~\bibnamefont
  {Wolschin}},\ }\bibfield  {title} {\bibinfo {title} {Solving a nonlinear
  analytical model for bosonic equilibration},\ }\href@noop {} {\bibfield
  {journal} {\bibinfo  {journal} {Physics Open}\ }\textbf {\bibinfo {volume}
  {2}},\ \bibinfo {pages} {100013} (\bibinfo {year} {2020})}\BibitemShut
  {NoStop}%
\bibitem [{\citenamefont {Davis}\ \emph
  {et~al.}(1995{\natexlab{b}})\citenamefont {Davis}, \citenamefont {Mewes},\
  and\ \citenamefont {Ketterle}}]{dmk95}%
  \BibitemOpen
  \bibfield  {author} {\bibinfo {author} {\bibfnamefont {K.~B.}\ \bibnamefont
  {Davis}}, \bibinfo {author} {\bibfnamefont {M.-O.}\ \bibnamefont {Mewes}},\
  and\ \bibinfo {author} {\bibfnamefont {W.}~\bibnamefont {Ketterle}},\
  }\bibfield  {title} {\bibinfo {title} {An analytical model for evaporative
  cooling of atoms},\ }\href@noop {} {\bibfield  {journal} {\bibinfo  {journal}
  {Appl. Phys. B}\ }\textbf {\bibinfo {volume} {60}},\ \bibinfo {pages} {155}
  (\bibinfo {year} {1995}{\natexlab{b}})}\BibitemShut {NoStop}%
\bibitem [{\citenamefont {Skeel}\ and\ \citenamefont
  {Berzins}(1990)}]{skeel1990method}%
  \BibitemOpen
  \bibfield  {author} {\bibinfo {author} {\bibfnamefont {R.~D.}\ \bibnamefont
  {Skeel}}\ and\ \bibinfo {author} {\bibfnamefont {M.}~\bibnamefont
  {Berzins}},\ }\bibfield  {title} {\bibinfo {title} {A method for the spatial
  discretization of parabolic equations in one space variable},\ }\href@noop {}
  {\bibfield  {journal} {\bibinfo  {journal} {SIAM J. Sci. Statist. Comput.}\
  }\textbf {\bibinfo {volume} {11}},\ \bibinfo {pages} {1} (\bibinfo {year}
  {1990})}\BibitemShut {NoStop}%
\bibitem [{\citenamefont {Svistunov}(1991)}]{svi91}%
  \BibitemOpen
  \bibfield  {author} {\bibinfo {author} {\bibfnamefont {B.~V.}\ \bibnamefont
  {Svistunov}},\ }\bibfield  {title} {\bibinfo {title} {Highly nonequilibrium
  {Bose} condensation in a weakly interacting gas},\ }\href@noop {} {\bibfield
  {journal} {\bibinfo  {journal} {J. Mosc. Phys. Soc.}\ }\textbf {\bibinfo
  {volume} {1}},\ \bibinfo {pages} {373} (\bibinfo {year} {1991})}\BibitemShut
  {NoStop}%
\bibitem [{\citenamefont {Kagan}\ and\ \citenamefont
  {Svistunov}(1997)}]{kas97}%
  \BibitemOpen
  \bibfield  {author} {\bibinfo {author} {\bibfnamefont {Y.}~\bibnamefont
  {Kagan}}\ and\ \bibinfo {author} {\bibfnamefont {B.~V.}\ \bibnamefont
  {Svistunov}},\ }\bibfield  {title} {\bibinfo {title} {Evolution of
  correlation properties and appearance of broken symmetry in the process of
  {Bose-Einstein} condensation},\ }\href@noop {} {\bibfield  {journal}
  {\bibinfo  {journal} {Phys. Rev. Lett.}\ }\textbf {\bibinfo {volume} {79}},\
  \bibinfo {pages} {3331} (\bibinfo {year} {1997})}\BibitemShut {NoStop}%
\bibitem [{\citenamefont {Pitaevskii}\ and\ \citenamefont
  {Stringari}(2003)}]{BookPitaevskii}%
  \BibitemOpen
  \bibfield  {author} {\bibinfo {author} {\bibfnamefont {L.}~\bibnamefont
  {Pitaevskii}}\ and\ \bibinfo {author} {\bibfnamefont {S.}~\bibnamefont
  {Stringari}},\ }\href@noop {} {\emph {\bibinfo {title} {Bose-Einstein
  condensation}}},\ International series of monographs on physics\ (\bibinfo
  {publisher} {Clarendon Press},\ \bibinfo {address} {Oxford},\ \bibinfo {year}
  {2003})\BibitemShut {NoStop}%
\bibitem [{\citenamefont {Kim}\ \emph {et~al.}(2018)\citenamefont {Kim},
  \citenamefont {Svidzinsky}, \citenamefont {Agrawal},\ and\ \citenamefont
  {Scully}}]{scul18}%
  \BibitemOpen
  \bibfield  {author} {\bibinfo {author} {\bibfnamefont {M.~B.}\ \bibnamefont
  {Kim}}, \bibinfo {author} {\bibfnamefont {A.}~\bibnamefont {Svidzinsky}},
  \bibinfo {author} {\bibfnamefont {G.~S.}\ \bibnamefont {Agrawal}},\ and\
  \bibinfo {author} {\bibfnamefont {M.~O.}\ \bibnamefont {Scully}},\ }\bibfield
   {title} {\bibinfo {title} {Entropy of the {{Bose}}-{{Einstein}}-condensate
  ground state: {{Correlation}} versus ground-state entropy},\ }\href@noop {}
  {\bibfield  {journal} {\bibinfo  {journal} {Phys. Rev. A}\ }\textbf {\bibinfo
  {volume} {97}},\ \bibinfo {pages} {013605} (\bibinfo {year}
  {2018})}\BibitemShut {NoStop}%
\bibitem [{\citenamefont {Yamamoto}\ and\ \citenamefont {Haus}(1986)}]{yam86}%
  \BibitemOpen
  \bibfield  {author} {\bibinfo {author} {\bibfnamefont {Y.}~\bibnamefont
  {Yamamoto}}\ and\ \bibinfo {author} {\bibfnamefont {H.~A.}\ \bibnamefont
  {Haus}},\ }\bibfield  {title} {\bibinfo {title} {Preparation, measurement and
  information capacity of optical quantum states},\ }\href@noop {} {\bibfield
  {journal} {\bibinfo  {journal} {Rev. Mod. Phys.}\ }\textbf {\bibinfo {volume}
  {58}},\ \bibinfo {pages} {1001} (\bibinfo {year} {1986})}\BibitemShut
  {NoStop}%
\end{thebibliography}%

\end{document}